\documentclass[a4paper, 11pt]{article}
\usepackage[sort]{natbib}
\usepackage[british]{babel}
\usepackage[pdftex]{graphicx}
\usepackage[svgnames,fixpdftex]{xcolor}
\usepackage{mathtools}
\usepackage[T1]{fontenc}
\usepackage[applemac]{inputenx}
\usepackage{float}
\usepackage{setspace}
\usepackage{fancyhdr}
\usepackage{xspace}
\usepackage{authblk}
\usepackage[scale=0.85]{geometry}
\usepackage{subcaption}
\usepackage[labelfont=bf,labelsep=period,font=small,textfont=it,width=0.9\textwidth, tableposition=top]{caption}
\usepackage[babel,kerning]{microtype}
\usepackage[scaled=0.95]{helvet,inconsolata}
\usepackage{booktabs}
\usepackage{titling}
\usepackage{xurl}
\usepackage[pdftex, pdfpagelabels, 
                 pdfpagelayout = TwoPageLeft, 
                 bookmarks,
                 bookmarksopen = false,
                 bookmarksnumbered = true,
                 breaklinks = true,
                 linktocpage,
				colorlinks=true, linkcolor=ao, urlcolor=NavyBlue, citecolor=blue, anchorcolor=blue,
				pdftitle={Kumaraswamy Bayesian},
				pdfsubject={Statistical inference and modelling},
				pdfauthor={ZS MAJ}]{hyperref}

\usepackage{rotating}
\usepackage{tikz}
\usetikzlibrary{positioning} 
\usetikzlibrary{arrows.meta}
\usepackage{adjustbox}
\usepackage{amssymb, amsmath}
\usepackage{upgreek}
\newcommand{\refi}[1]{Figure~\ref{fig:#1}}
\newcommand{\reta}[1]{Table~\ref{tab:#1}}
\DeclarePairedDelimiterX{\cb}[1]\{\}{#1}
\DeclarePairedDelimiterX{\rb}[1](){#1}
\DeclarePairedDelimiterX{\sqb}[1][]{#1}

\newcommand{\sqbra}[1]{\sqb*{#1}\xspace}
\newcommand{\paren}[1]{\rb*{#1}\xspace}
\newcommand{\cubra}[1]{\cb*{#1}\xspace}

 \DeclareMathOperator{\var}{\mathbb V} \DeclareMathOperator{\E}{\mathbb E}  
\DeclareMathOperator{\med}{Med}

\newcommand{\Real}{\ensuremath{\mathbb R\xspace}}
\newcommand{\bl}[1] {\ensuremath{\boldsymbol{#1}}\xspace}
\newcommand{\samp}[3][1]{\ensuremath{\cubra{#3_{#1},\dots,#3_{#2}}}\xspace}

\newcommand{\g}{\, \left. \right|}

\newcommand{\ie}{\emph{i.e.}\xspace}
\newcommand{\eg}{\emph{e.g.}\xspace}
\newcommand{\ia}{\emph{e.g.}\xspace}
\newcommand{\kum}{\ensuremath{\mathrm K}\xspace}
\newcommand{\ekum}{\ensuremath{\mathrm{eK}}\xspace}
           
\DeclarePairedDelimiterX\pafun[2](){#1\nonscript\>\delimsize\vert\nonscript\> \mathopen{}#2}
\DeclarePairedDelimiterX\papro[2][]{#1\nonscript\>\delimsize\vert\nonscript\> \mathopen{}#2}
\newcommand\mo[3][f]{\ensuremath{#1\pafun{#2}{#3}}}
\newcommand{\pri}[1][\theta]{\ensuremath{\uppi\paren{#1}}\xspace}
\newcommand{\mop}[3][\uppi]{\mo[#1]{#2}{#3}\xspace}

\DeclarePairedDelimiterX\paden[3](){\nonscript #1 \,\delimsize\vert \,\nonscript \mathopen{}#2,#3}
\DeclarePairedDelimiterX\pastu[4](){\nonscript #1\,\delimsize\vert\,\nonscript \mathopen{}#2,#3,#4}
\DeclarePairedDelimiterX\paek[4](){\nonscript #1\,\delimsize\vert\,\nonscript \mathopen{}#2,#3,#4, l, u}

\newcommand{\ga}[3][x]{\ensuremath{\mathrm{Ga} \paden*{#1}{#2}{#3}}\xspace}

\newcommand{\be}[3][x]{\ensuremath{\mathrm{Be} \paden*{#1}{#2}{#3}}\xspace}

\newcommand{\un}[3][x]{\ensuremath{\mathrm{Un} \paden*{#1}{#2}{#3}}\xspace}

\newcommand{\ek}[4][x]{\ensuremath{\mathrm{eK} \paek*{#1}{#2}{#3}{#4}}\xspace}
\newcommand{\ku}[3][x]{\ensuremath{\mathrm{K} \paden*{#1}{#2}{#3}}\xspace}
\newcommand{\bsc}[5][x]{\ensuremath{\mathrm{Be}_{(#4, #5)} \paden*{#1}{#2}{#3}}\xspace}
\definecolor{ao}{rgb}{0.4, 0.0, 0.9}

\graphicspath{{../Figures/}}

\begin{document}
\title{\textbf{Bayesian inference on the exponentiated Kumaraswamy distribution with applications} \\[-1em]}
\author{Ziqian Sun}
\author{Miguel A. Ju‡rez}

\affil{School of Mathematical and Physical Sciences \\ The University of Sheffield, UK}
\date{}

\maketitle

\begin{abstract}
	We discuss Bayesian estimation of the exponentiated Kumaraswamy distribution, based on a location-scale-shape parameterisation.  The parameterisation facilitates prior elicitation and interpretation of results, but potentially entails identifiability issues that are addressed through a hierarchical weakly informative prior setting. Our HMC implementation enables off-the-shelf utility for practitioners, and is illustrated on synthetic and real data. 
\end{abstract}

\section{Bounded outcome scores}  \label{sec:bos}

Analysis of observations bounded within a finite interval, often referred to as bounded outcome scores (BOS), is pervasive in several fields \citep[see][\ia]{shad, hu, Verkuilen, crema}, prompting a wide range of approaches for dealing with the various types of BOS. Most often through the use of transformations either to the unit range, or an unbounded range so well known models can be applied  \citep[\eg][]{hunger2012longitudinal_beta_hrql, parker2024BB}. Originating in hydrology research, the Kumaraswamy (\kum) distribution \citep{kumaraswamy1980} is a flexible and bounded-support distribution that has been used for analysing BOS in several fields \citep{Garg2008, jones2009kbeta, Lemonte2011KumPE, lemonte2013EK, mitnik2013median, mitnik2013new, Kohansal2019}. It is related to the generalised Beta distribution family  \citep{nadarajah2008discussion}, and is a special case of the McDonald's generalised Beta distribution of the first kind \citep{mcdonald1984gb1}. \citet{jones2009kbeta} provides a thorough review on the advantages and disadvantages of the \kum distribution and compares it with the Beta distribution from multiple perspectives in terms of basic distributional properties and modelling tractability, among others. \citet{Dey2018} and  \citet{Wang17} provide and compare several frequentist estimation methods, while \citet{shi, dey2016} and \citet{huanmin} propose Bayesian alternatives.  

In this paper, we provide a fully Bayesian approach to inference for continuous BOS data with an extension of the \kum distribution, the exponentiated Kumaraswamy (\ekum) distribution \citep{lemonte2013EK}.  Section \ref{sec:ek} formalises the approach and introduces two location-scale parameterisations that facilitate prior setting and elicitation. The Bayesian model is formalised in Section~\ref{sub:bayes}, with Section~\ref{sub:implementation} detailing the inference implementation. Application to two real data sets is illustrated in Section~\ref{sub:apps}.

\section{The exponentiated Kumaraswamy distribution} \label{sec:ek}

Formally, we say that the observable $x$, follows the exponentiated Kumaraswamy distribution, $\ek\sigma\kappa\gamma$, with support $\mathcal X = (l, u)$, and parameters $\paren{\sigma, \kappa, \gamma} \in \Real^3_+$, if its CDF is given by,
\begin{equation}
	F_{\ekum}(x \g \sigma, \kappa, \gamma, l, u) = \cubra{ 1 - \sqbra{1 - \paren{ \frac{x - l}{u - l}}^{\sigma}}^{\kappa}}^{\gamma}; \quad l < x < u.
\label{eq:EK_cdf}
\end{equation}
This was introduced by \citet{lemonte2013EK}, based on the \kum distribution, by adding a shape parameter, $\gamma$, for increased flexibility. Due to its closed analytical expression, the $\ekum$ can be parameterised as a location-scale-shape model \citep{mitnik2013median, lemonte2013EK}, with shape parameter $\gamma$; location , $\omega$, satisfying,
\begin{equation}
	\cubra{ 1 - \sqbra{1 - \paren{\frac{\omega - l}{u - l}}^\sigma}^\kappa}^\gamma = 0.5; \label{eq:EK_median}
\end{equation}
and scale parameter either
\begin{equation}
\kappa(\omega, \sigma, \gamma) = \frac{\log \left( 1 - 0.5^{1/\gamma} \right) }{\log \left( 1 - \left( \frac{\omega - l}{u - l} \right)^\sigma \right)
}, \quad \text{or} \quad
\sigma(\omega, \kappa, \gamma) = \frac{\log \!\left( 1 - \left( 1 - 0.5^{1/\gamma} \right)^{1/\kappa} \right)}{\log \!\left( \frac{\omega - l}{u - l} \right)}. \label{eq:relat_ka_si_ga}
\end{equation}

\begin{figure}[!ht]
    \centering
        \includegraphics[width=\linewidth]{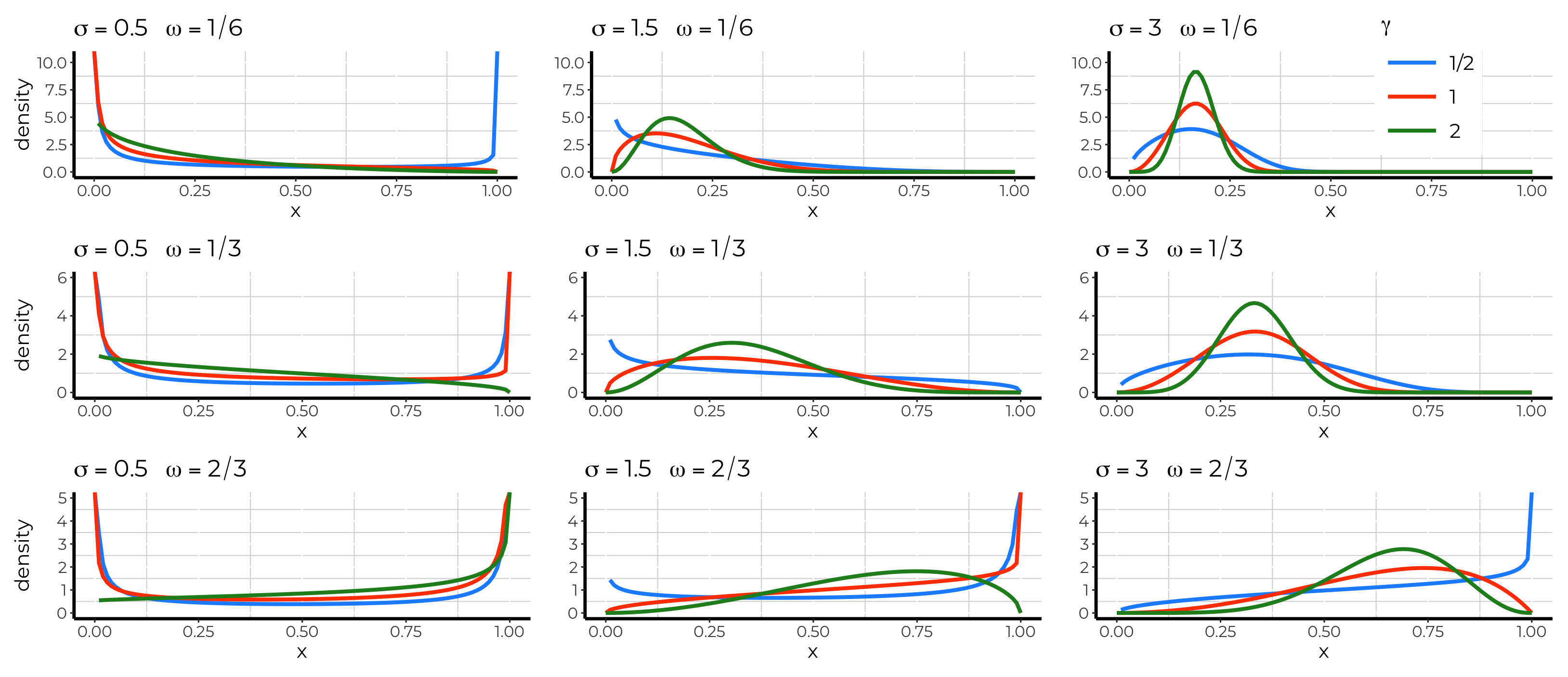}
    \caption{The \ekum distribution bounded in $(0, 1)$, with the location-scale-shape parameterisation. Each column has a different scale parameter $\sigma = 	0.5, 1.5, 3$, each row has a different location parameter $\omega = 1/6, 1/3, 2/3$. within each cell a different shape parameter $\gamma = 1/2, 1, 2$ in blue, red and green, respectively. Note that larger values of the shape parameter, $\gamma$, result on a more regular distribution, while smaller values of the scale, $\sigma$, result on larger mass close to either boundary, or both.}   \label{fig:ek_shape}
\end{figure}

Figure \ref{fig:ek_shape} illustrates the flexibility of the \ekum distribution, with the location-scale-shape parameterisation. The shape of the distribution can be almost symmetric (either bell- or U-shaped) or left- or right-skewed. Notice that, regardless of the location, the distribution accumulates mass at either or both boundaries for small values of the scale, with larger values of the shape, $\gamma$, smoothing the tails, as shown mainly in the leftmost column of the figure.

\subsection{The Kumaraswamy Distribution} \label{sec:K_properties}

A particular case of the \ekum distribution when $\gamma = 1$, the Kumaraswamy (\kum) distribution has been extensively used in practice, and compared with alternative models \citep[see \eg][and references therein]{huanmin, abo, nadar, hassan}. Here we provide a summary of some of its properties and relation to the Beta family of distributions.  Formally, we say that the observable, $x$, with support in $(l, u)$, follows \kum distribution,  $ X \thicksim \ku\sigma\kappa$, if its PDF is given by
\begin{equation}
	\ku\sigma\kappa = 	\frac{\sigma \kappa}{u - l}  \paren{\frac{x - l}{u - l}}^{\sigma - 1} \sqbra{1 - \paren{\frac{x - l}{u - l}}^{\sigma}}^{\kappa - 1}; 	\quad l < x < u. \label{eq:K_pdf_cdf_inv}		
\end{equation}
with mean
\begin{gather*}
\E[X] = l + (u - l)\> \kappa \> \mathrm B \paren{1 + \frac{1}{\sigma}, \kappa},
\shortintertext{and variance}	
\var[X] = (u - l)^2 \cubra{\kappa \> \mathrm B\paren{1 + \frac{2}{\sigma}, \kappa} -  \sqbra{\kappa \> \mathrm B \paren{1 + \frac{1}{\sigma}, \kappa}}^2} ,
\end{gather*}
where $\mathrm B( a, b)$ is the Beta function \citep{mitnik2013new}. From the quantile function, the median of the distribution is,
\[
\med[X] = \omega(\sigma, \kappa) = l + (u - l) \, \paren{1 - 0.5^{1/\kappa}}^{1/\sigma}.
\]
For fixed $(l, u)$, we say that $z = (x - l) / (u - l)$ follows the standard \kum with pdf,
\[
	\ku[z]\sigma\kappa = 	\sigma \kappa \> z^{\sigma - 1} \paren{1 - z^\sigma}^{\kappa - 1}, 	\quad 0 < z < 1.
\]

\subsection{Related distributions}\label{sec:K_vs_beta}

The \kum and Beta distribution are related through the Generalised Beta distribution of the First Kind ($\mathrm{GB}$) \citep{mcdonald1984gb1},
\begin{equation}
\mathrm{GB}(x \g \sigma, \alpha, \kappa, u) = \frac{\sigma \> x^{\sigma \alpha - 1} \paren{ 1 - \paren{\frac{x}{u}}^{\sigma}}^{\kappa - 1}}
{u^{\sigma \alpha} \> \mathrm{B}(\alpha, \kappa)}, \quad x\in(0, u). \label{eq:pdf_gb1}
\end{equation}
When $\sigma = u = 1$, \eqref{eq:pdf_gb1} reduces to a \be\alpha\kappa.

When $\alpha = 1$, 
\begin{equation}
\mathrm{GB}(x \g \sigma, \kappa, u) = \frac{\sigma \kappa}{u} \, \paren{\frac{x}{u}}^{\sigma - 1} \paren{1 - \paren{ \frac{x}{u}}^{\sigma}}^{\kappa - 1}, \quad x \in (0, u), \label{eq:GB1_p1}
\end{equation}
is a \ku\sigma\kappa with support on $\mathcal X = (0, u)$. Letting $y = l + \frac{u - l}{u}x$, for $-\infty < l < u < \infty$, yields
\begin{equation*}
	\mo{y}{\sigma, \kappa} = \frac{\sigma \kappa}{u - l} \> \paren{\frac{y - l}{u - l}}^{\sigma - 1}
   \sqbra{1 - \paren{\frac{y - l}{u - l}}^{\sigma}}^{\kappa - 1},    \qquad l < y < u, %\label{eq:pdf_K_lu}
\end{equation*}
a \ku[y]\sigma\kappa, with support, $\mathcal Y = (l, u)$ as in \eqref{eq:K_pdf_cdf_inv}. Note that the $\mathrm{GB}$ is in the Exponential family if and only if $\sigma =1$, hence the \kum distribution is outside the family.  We provide a graphical summary in \refi{gb1-special-cases} and refer the reader to \citet{shi} for a more thorough filiation of the Kumaraswamy distribution. 

\begin{figure}[!ht]
	% \begin{adjustbox}{max width=\textwidth}
	\begin{tikzpicture}[scale=0.17, transform shape,
		font=\Huge,
	  node distance=3cm and 3cm,
	  box/.style={
	    draw=none, rounded corners, align=center,
	    text width=18cm, minimum height=7.5cm
	  },
	  arrow/.style={-Stealth, thick}
	]

	% GB1 at the top
	\node[box] (GB1) {
	  \textbf{Generalised Beta Distribution of the First Kind (GB)}\\[1ex]
	   \[
	   \mo[\mathrm{GB}]{x}{\sigma, u, \alpha, \kappa} = \frac{\sigma x^{\sigma p - 1} \paren{1 - (x / u)^\sigma}^{\kappa - 1}}{u^{\sigma \alpha} \mathrm B(\alpha, \kappa)}, \quad x \in (0, u) \> \sigma, u, \alpha, \kappa > 0
	   \].
	};

	\node[box, below left=of GB1] (Kum) {
	  \textbf{Special Case 1: $\kum$ distribution}\\[1ex]
	  \[\ku \sigma\kappa = \frac{\sigma \kappa}{u - l} 
	   \paren{ \frac{y - l}{u - l}}^{\sigma - 1}
	   \sqbra{ 1 - \paren{\frac{y - l}{u - l}}^{\sigma}}^{\kappa - 1}, 
	   \quad  x \in (l, u)\].
	};

	\node[box, below left=of Kum] (Kum-unit) {
	  \textbf{Standard \kum distribution}\\[1ex]
	  \[\ku[z]\sigma \kappa) = \sigma \kappa x^{\sigma - 1}(1 - x^\sigma)^{\kappa - 1}, \quad x \in (0, 1)\]
	};

	\node[box, below right=of Kum] (Exponentiated-Kum) {
	  \textbf{\ekum distribution}\\[1ex]
	  \[\ek\sigma\kappa\gamma = 
	\frac{\gamma \sigma \kappa}{u - l}
	\left( \frac{x - l}{u - l} \right)^{\sigma - 1}
	\left[ 1 - \left( \frac{x - l}{u - l} \right)^\sigma \right]^{\kappa - 1}
	\left\{ 1 - \left[ 1 - \left( \frac{x - l}{u - l} \right)^\sigma \right]^\kappa \right\}^{\gamma - 1}, \quad x \in (l, u)\]
	};

	\node[box, below right=of GB1] (Beta) {
	  \textbf{Special Case 2: Beta Distribution}\\[1ex]
	   \[
		\be \alpha\kappa = 
		  \frac{1}{\mathrm B(\alpha, \kappa)} x^{\alpha - 1} (1 - x)^{\kappa - 1}, \quad x \in (0, 1)
	   \]
	};

	% Arrows with labels
	\draw[arrow] (GB1) -- node[midway, left, xshift=-25pt, yshift=20pt] {$\alpha=1$, $y=l+\frac{u-l}{u}x$} (Kum);
	\draw[arrow] (GB1) -- node[midway, right, xshift=25pt] { $\sigma=1$, $u=1$} (Beta);
	\draw[arrow] (Kum) -- node[midway, left,yshift=15pt] { $l=0$, $u=1$} (Kum-unit);
	\draw[arrow, bend left=15] (Kum) to node[midway, right, xshift=20pt] {shape parameter, $\gamma$} (Exponentiated-Kum);
	\draw[arrow, bend right=30] (GB1) to node[midway, below, yshift=53pt] { $\alpha=1$, $u=1$} (Kum-unit);
	\draw[arrow, bend left=15] (Exponentiated-Kum) to node[midway, left, xshift=-40pt] { $\gamma=1$} (Kum);

	\end{tikzpicture}
	% \end{adjustbox}
	\caption{Relationship between the Generalised Beta distribution of the first kind ($\mathrm{GB}$), the Kumaraswamy (\kum) distribution, the Exponentiated Kumaraswamy (\ekum) distribution, and the Beta distribution. The diagram shows that \kum and Beta are special cases of $\mathrm{GB}$. \kum is a simplification of the \ekum by letting the shape parameter $\gamma=1$.} \label{fig:gb1-special-cases}
\end{figure}
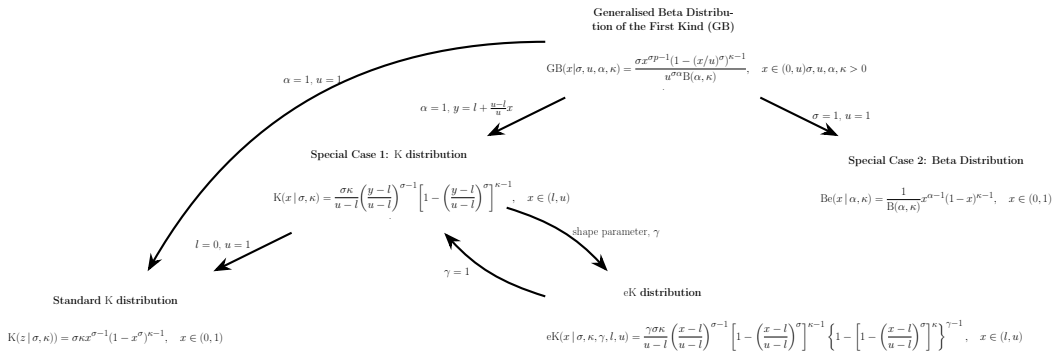

\section{Bayesian inference} \label{sub:bayes}

We assume the observations are a random sample $\bl x = \samp nx$, where $x_i \thicksim \ek[x_i]{\omega}{\theta}{\gamma}$ with the location, $\omega$, as in \eqref{eq:EK_median} and the scale, $\theta$, either $\sigma$ or $\kappa$, as defined by \eqref{eq:relat_ka_si_ga}. The key advantages from this are the direct interpretation of the median as a location parameter, making it relatively straightforward to set weak prior, \ie $\pri[\omega] = \un[\omega]lu$, or to elicit an informative one. Two relatively natural options are a \ku[\omega]{w}{s} with support in $(l, u)$, or a Beta distribution scaled to $(l, u)$
\begin{equation}
	 \bsc[\omega]ablu = \frac{1}{(u - l)} \frac 1{\mathrm B(a,b)} \paren{\frac{\omega - l}{u - l}}^{a - 1} \paren{1 - \frac{\omega - l}{u - l}}^{b - 1}; \quad l < \omega < u. \label{eq:rebeta}
\end{equation}

It would seem almost natural to set a Gamma prior for either scale parameter \citep[as in][for example]{nadar}, but the parameterisation carries a potential identifiability issue, as illustrated in the left hand side of \refi{sens}.  Further, it is apparent that the scales of each parameter are quite different, hence setting a Gamma distribution for either would call for appropriate specific prior settings.  Instead, we propose a hierarchical prior, $\pri = \ga[\theta]{a}\psi \ga[\psi]bc$, to induce flexibility and provide a unified modelling approach, regardless of the selected parameterisation, $\theta = \sigma$ or  $\kappa$.  This distribution has a finite variance if $b>2$, and a mode at $c (a -1) /(b+1)$ if $a>1$. In practice, we recommend $a=4.5$, $b=2.5$, to allow for a large, finite variance and to let $c$ control the mode.

As a weakly informative prior for the shape parameter, $\gamma$, we suggest a \kum distribution with mode at 1 and median at 3, $\ku[\gamma]{1.1}{4}$, bounded within $(0.4, 20)$, as illustrated on the left hand side of \refi{pripre}.   This setting centres the prior at the simpler model with a reasonably large variance. The key argument for restricting the parameter space is that this is consistent with most applications and helps the sampler to converge by discarding regions where the likelihood varies sharply close to the bounds of the sampling space. As illustrated in the right had side of \refi{sens}, for small values of $\gamma$, the \ekum accumulates more mass close to either boundary, and its shape regularises as $\gamma$ increases.  In practice, we argue this restriction is warranted, unless there is compelling empirical evidence in favour of a sharp U, L or J shape. 

\begin{figure}
\centering
\subfloat[Location-scale-shape parameterisation of the \ekum distribution.  The relationship between $\kappa$ and $\sigma$ for fixed values of the median $\omega$ and shape, $\gamma$.  It is apparent that wide ranges of large values of $\kappa$ yield virtually the same $\omega$ for a small value of $\sigma$.]{%
\resizebox*{7cm}{!}{\includegraphics[keepaspectratio, width = 0.97\textwidth]{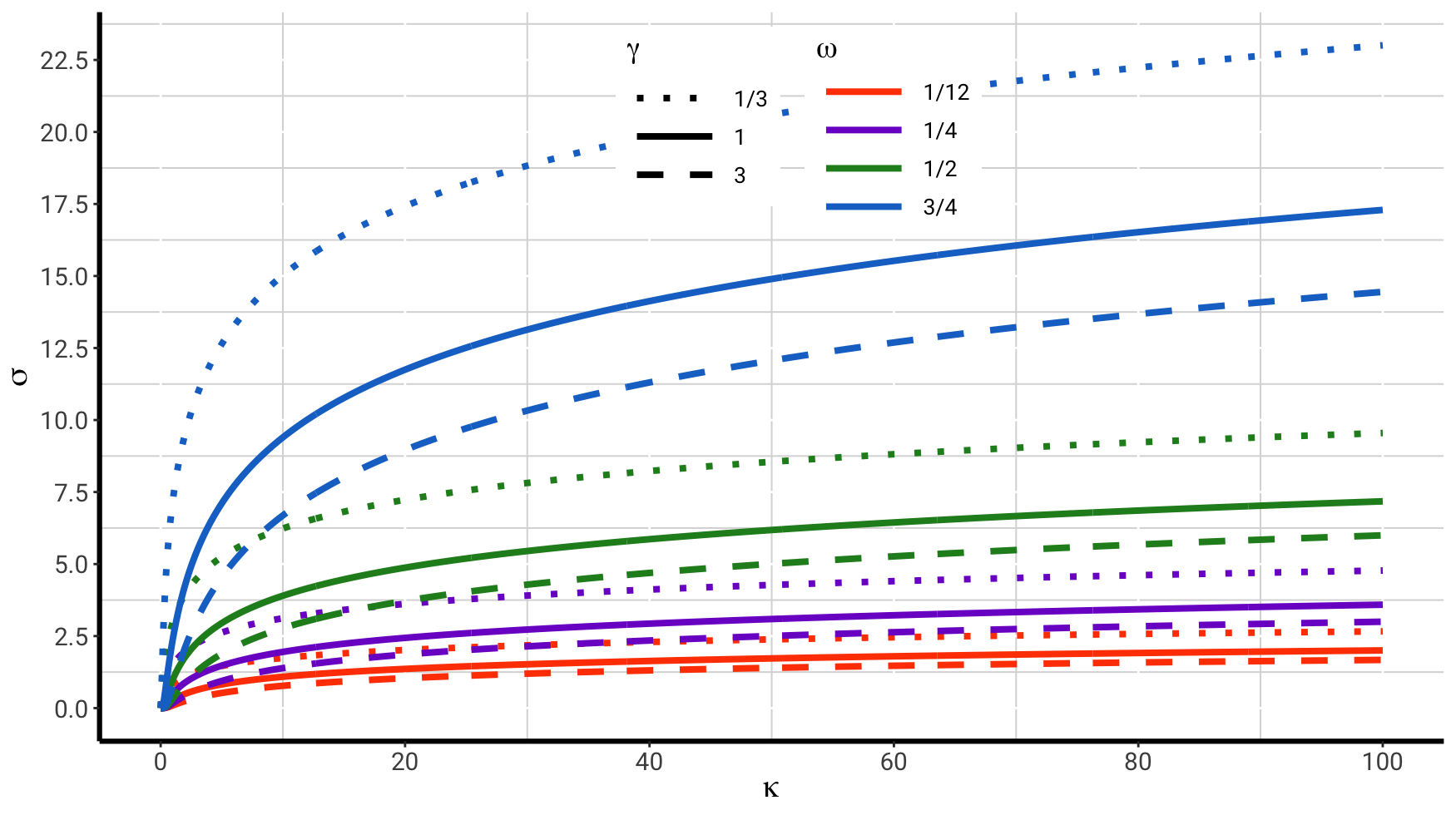}}}  \hspace{5pt}
\subfloat[PDF of the \ekum for a fixed median, $\omega=1/3$, and varying values of the scale, $\sigma$, and shape, $\gamma$, parameters.  Note how the distribution accumulates mass closer to the boundaries when $\gamma$ is relatively small and how its shape regularises as $\gamma$ increases.]{%
\resizebox*{7cm}{!}{\includegraphics[keepaspectratio, width = 0.97\textwidth]{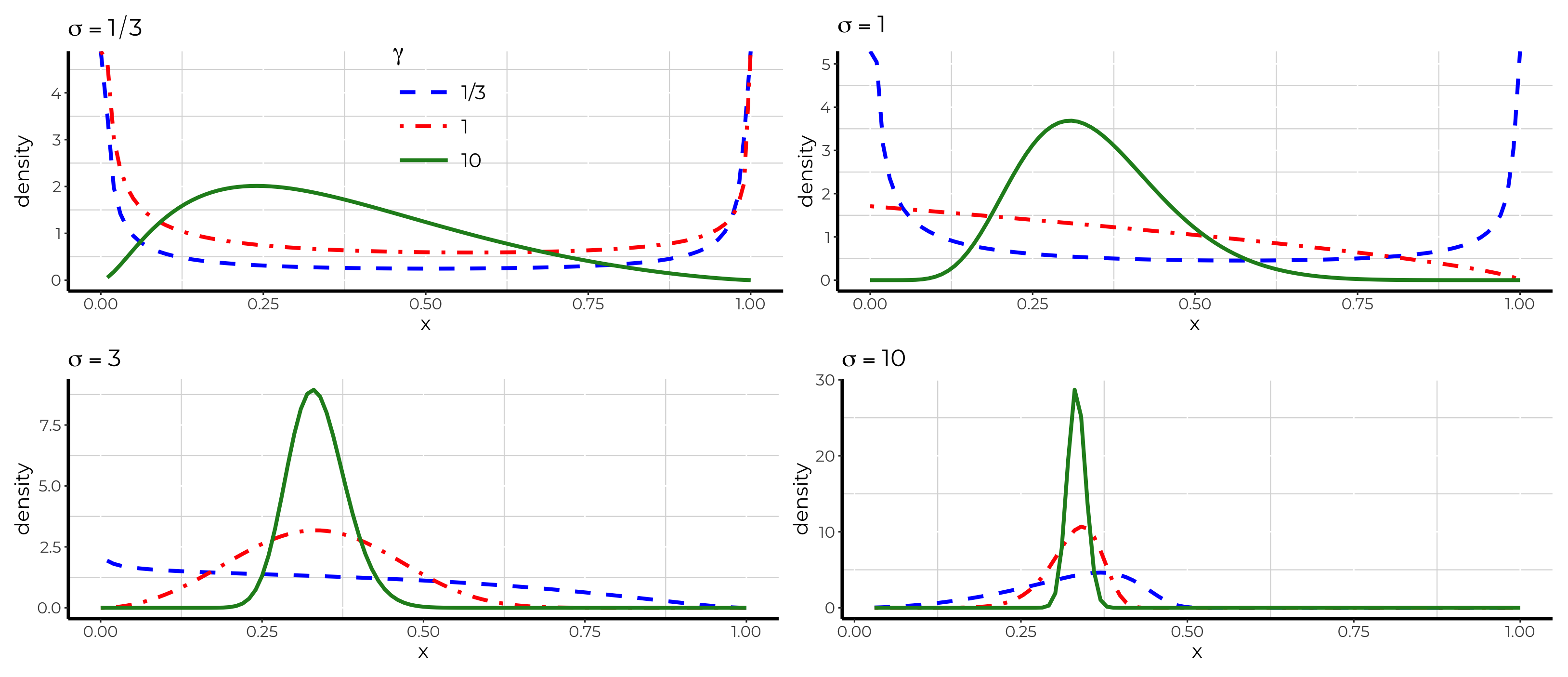}}}	
\caption{The \ekum probability density function.  The left hand side illustrates the relationship between the scale parameters for fixed values of the location and shape; note the potential identifiability issues for a fixed median and relatively low values of $\sigma$. The right hand side highlights the argument for restricting the parameter space for the shape, $\gamma$, unless there is strong empirical evidence of a U, L or J shape distribution.} \label{fig:sens}
\end{figure}

% \begin{figure}[!ht]
% 	\centering
% 		\begin{subfloat}[Location-scale-shape parameterisation of the \ekum distribution.  The relationship between $\kappa$ and $\sigma$ for fixed values of the median $\omega$ and shape, $\gamma$.  It is apparent that wide ranges of large values of $\kappa$ yield virtually the same $\omega$ for a small value of $\sigma$.]{\resizebox*{5cm}{!} \includegraphics[keepaspectratio, width = 0.97\textwidth]{kapasigma}}\label{fig:kasi} \hspace{5pt}
% 		\end{subfloat}
% 		\begin{subfloat}[PDF of the \ekum for a fixed median, $\omega=1/3$, and varying values of the scale, $\sigma$, and shape, $\gamma$, parameters.  Note how the distribution accumulates mass closer to the boundaries when $\gamma$ is relatively small and how its shape regularises as $\gamma$ increases.]{\resizebox*{5cm}{!} \includegraphics[keepaspectratio, width = 0.97\textwidth]{ek_gamasens}}	\label{fig:gamsens}
% 		\end{subfloat}
% 	\caption{The \ekum probability density function.  The left hand side illustrates the relationship between the scale parameters for fixed values of the location and shape; note the potential identifiability issues for a fixed median and relatively low values of $\sigma$. The right hand side highlights the argument for restricting the parameter space for the shape, $\gamma$, unless there is strong empirical evidence of a U, L or J shape distribution.}	\label{fig:sens}
% \end{figure}

Hence, we complete the Bayesian model assuming either $x_i \thicksim \mo[\mathrm{EK}_\sigma]{x_i}{\omega, \sigma, \gamma, l, u}$, or $x_i \thicksim \mo[\mathrm{EK}_\kappa]{x_i}{\omega, \kappa, \gamma, l, u}$, independent for $i= 1, \dots, n$, with the weakly informative prior setting 
\begin{equation}
	 \label{pri_weak}
	\begin{aligned}
		\pri[\omega] &= \un[\omega]lu \\ 
		\mop\theta{a, b, c} &= \ga[\theta]{a}{\psi} ~ \ga[\psi]bc; \quad \theta = \sigma, \kappa \\
		\pri[\gamma] &= \ku[\gamma]sk; \quad 0.4 < \gamma <20. 
	\end{aligned}
\end{equation}
or the informative one
\begin{equation}
	 \label{pri_info}
	\begin{aligned}
		\pri[\omega] &= \bsc[\omega]ablu  \\ 
		\mop\theta{a, b, c} &= \ga[\theta]{a}{\psi} ~ \ga[\psi]bc; \quad \theta = \sigma, \kappa \\
		\pri[\gamma] &= \ku[\gamma]sk;  \quad 0.4 < \gamma < 20. 
	\end{aligned}
\end{equation}

The right hand side of \refi{pripre} illustrates the prior predictive distribution from the \eqref{pri_weak}, with $c=1$.  It is apparent that this prior setting give rise to a prior predictive distribution that is almost flat over the sample space, with some bumps close to its boundaries.  The figure also shows the prior predictive from a single-level Gamma prior, $\pri[\gamma] = \ga[\gamma]{2}{0.6}$; this prior setting is chosen to allow the prior to have a mode away from zero, with a large variance. Even though both prior predictive distributions are very similar, in practice the single layer prior typically shrinks the posterior strongly towards the prior mode, while the power tails from the hierarchical setting avoid such aggressive shrinkage.

\begin{figure}[!ht]
	\centering
		\includegraphics[keepaspectratio, width = \textwidth]{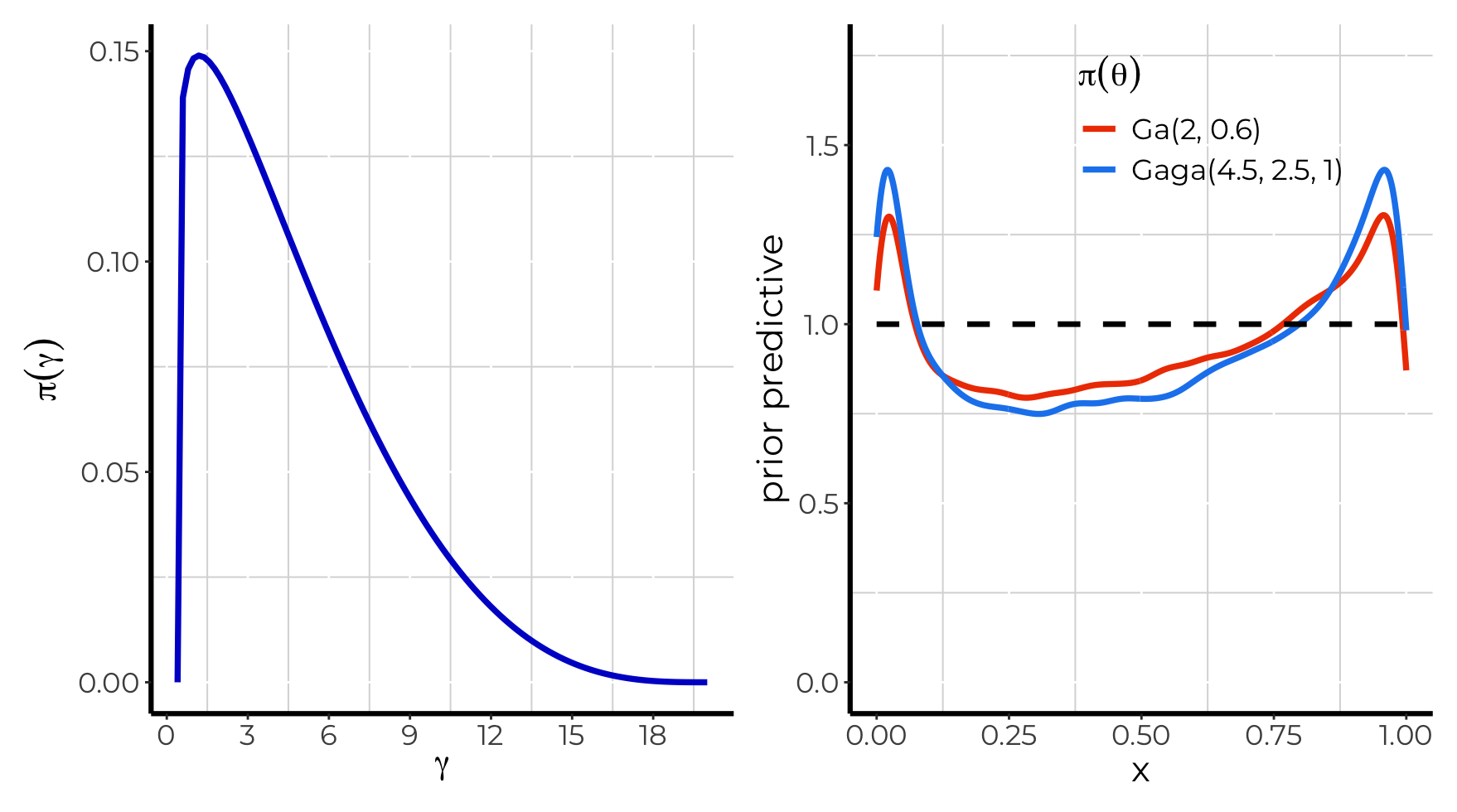}
	\caption{Left hand side shows the prior of the shape parameter, $\gamma$ and the right hand side the implied prior predictive distribution from the weakly informative prior setting in blue dashed, and the one from the one-level Gamma prior in orange solid. The grey line is a uniform prior over the sample space.}	\label{fig:pripre}
\end{figure}

\section{Implementation} \label{sub:implementation}

We evaluate our approach on a series of synthetic data sets.  In this study we generate samples of small, moderate, large and substantial sizes to gauge the influence of the weakly informative prior setting, with different combinations of parameters to cover a wide range of scenarios. We simulate from the standard \kum and, to avoid cluttering, report only our main findings.  The data are fitted through HMC in R \citep{R} using Stan \citep{stan}, the code and full sensitivity analysis are available from the corresponding author upon request.

We ran the sampler with the weakly informative prior with $c=1$ over the 240 instances arising from the all possible parameter combinations when fixing
\[
	\omega = 1/6, 1/3, 2/3; \quad \kappa = 3/4, 2, 10, 50, 100; \quad \gamma = 3/4, 1, 3, 5;
\]
for sample sizes $n= 25, 100, 500, 2000$.  As expected, the Stan sampler needed calibration of the adaptive control settings for those instances with the smaller values of either location, shape or scale parameters, regardless of the sample size.  In those cases, the posterior distribution of the shape was determined by the prior in all but the largest sample size (see \refi{simxtrm}); the posterior distribution of the median covered its actual value and the predictive distribution resembled the empirical distribution closely, regardless.  \enlargethispage*{1ex}

\begin{figure}[!ht]
	\centering
		\includegraphics[keepaspectratio, width=\textwidth]{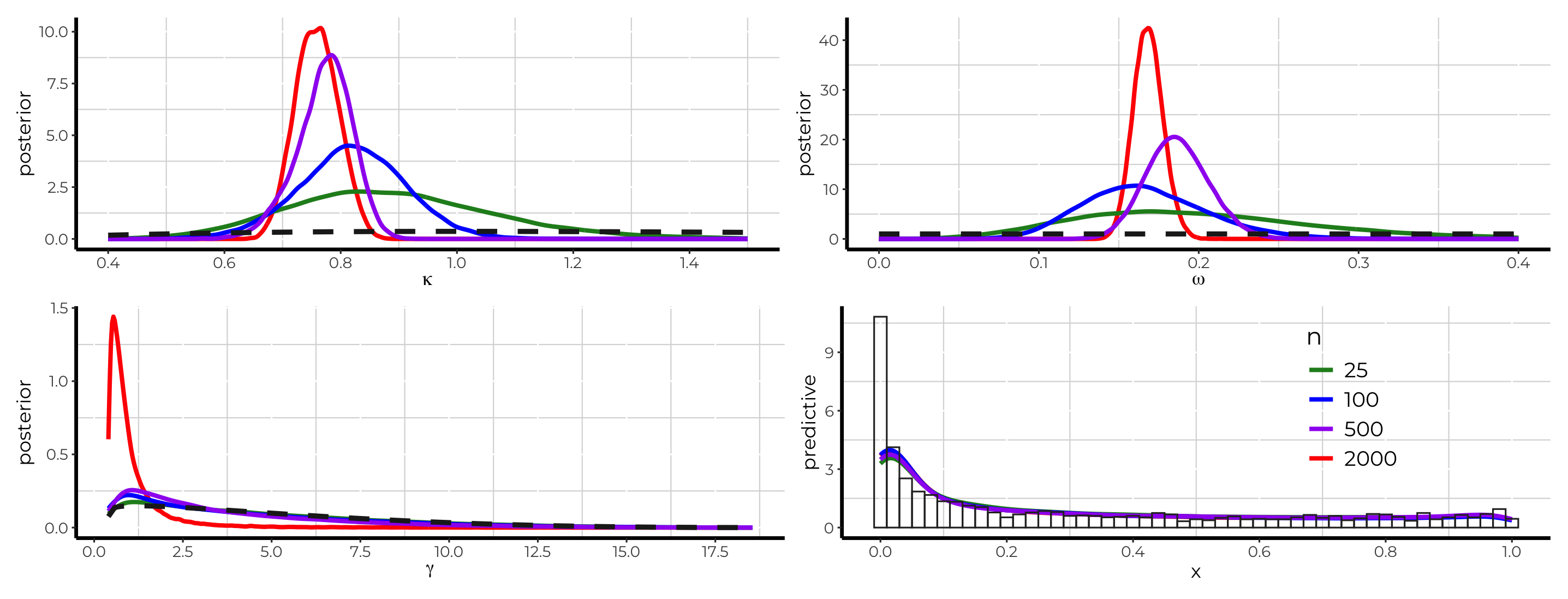}
	\caption{Posterior and predictive distributions from instances with $\omega = 1/6$, $\kappa=3/4$, $\gamma=3/4$, and sample sizes $n = 25, 100, 500, 2000$; priors depicted with the dashed grey lines.  Note the wide posterior ranges for the shape parameter, regardless of the sample size.  Note the posterior of the shape, $\gamma$, follows the prior for all but the largest sample size, while the posterior of the scale and location are increasingly accurate and precise.  The predictive distribution recovers the empirical distribution in any case---histogram of the largest data set shown.}
	\label{fig:simxtrm}
\end{figure}

For instances yielding a more regular shape, the samplers did not need specific control adjustments and ran substantially faster; the posterior distributions covered the underlying parameters well and the predictive distributions recovered the empirical distribution closely, as summarised in \refi{simregs}.

\begin{figure}[!ht]
	\centering
		\includegraphics[keepaspectratio, width=\textwidth]{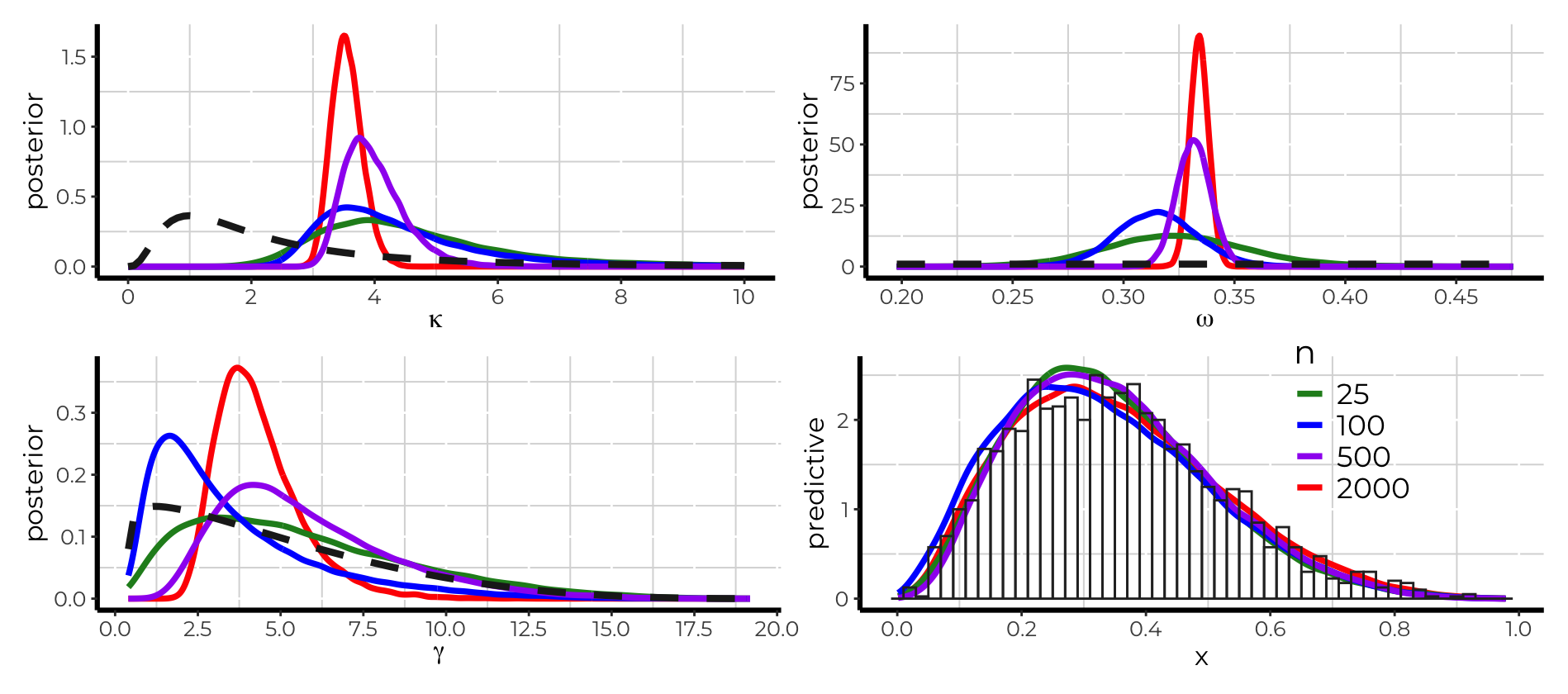}
	\caption{Typical behaviour of the Bayesian learning process for instances with the shape and scale parameters well within their parameter spaces.  In the illustration, $\omega = 1/3$, $\kappa = 4$, $\gamma=3$, and sample size $n=25, 100, 500, 2000$.  Priors plotted with the dark, dashed lines.  As the sample size increases, the posterior distribution location converge to the corresponding underlying values, with increasing precision.  The predictive distribution recovers the empirical distribution, regardless of the sample size---histogram of the larger data set shown.}
	\label{fig:simregs}
\end{figure}

\section{Applications} \label{sub:apps}

To illustrate our approach we analyse two datasets. Each was fitted using both prior specifications, \eqref{pri_weak} and \eqref{pri_info}; we show the results from the informative setting as they are virtually undistinguishable from the alternative prior. The first set, dating back to \citet{bartlett}, was analysed by \citet{jorgensen} using an inverse Gaussian distribution; more recently, \citet{Lemonte} fitted a log-exponentiated Kumaraswamy distribution.  The data we use \citep[available from][]{GeneralizedHyperbolicPackage}, illustrated on the left hand side of \refi{dats}, comprise intervals between 129 successive vehicles pass a point on a road, measured in seconds. The observed range is (0.2, 125.3), with IQR 14.57, sample mean 15.81, median 5.85, standard deviation 23.7, Pearson skewness 2.51, and kurtosis 9.66. 

\begin{figure}[!ht]
	\centering
		\includegraphics[keepaspectratio, width = \textwidth]{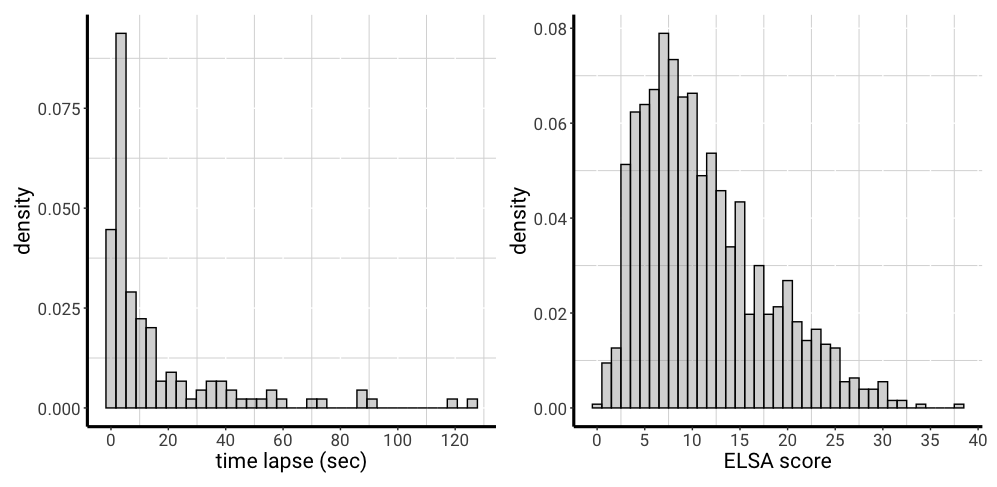}
	\caption{Histogram of the two datasets used for illustration. The left hand side depicts 128 time lapses between consecutive vehicles in a particular stretch of road, taken from \citet{GeneralizedHyperbolicPackage}.  The right hand side summarises 1267 frailty scores taken at baseline from the longitudinal CARE75+ study \citep{Heavene}.}	\label{fig:dats}
\end{figure}

We set the support of the distribution to $(l, u) = (0.1, 130)$, and centre the informative prior for the location at 6. So,
\begin{align}
	\pri[\omega] &= \bsc[\omega]{0.1}{2}{0.1}{130} \\ \notag
	\pri[\theta] &= \ga[\theta]{4.5}{\psi} \, \ga[\psi]{2.5}{1}, \quad \theta = \sigma, \kappa \\ \label{pri_traf}
	\pri[\gamma] &= \ku[\gamma]{1.1}{4}; \quad 0.4 < \gamma < 20. \notag
\end{align}

The sampler was run for each parameterisation, $\theta$, with 3 chains in parallel, burning-in the first 5k draws and recording the next 20k, ending up with samples 45k for inference. The average runtime was 6 seconds on an Apple silicon M2 with 16GB of RAM.  Posterior distributions and the posterior predictive are depicted in \refi{pos_traf}. 

\begin{figure}[!ht]
	\centering
		\includegraphics[keepaspectratio, width = \textwidth]{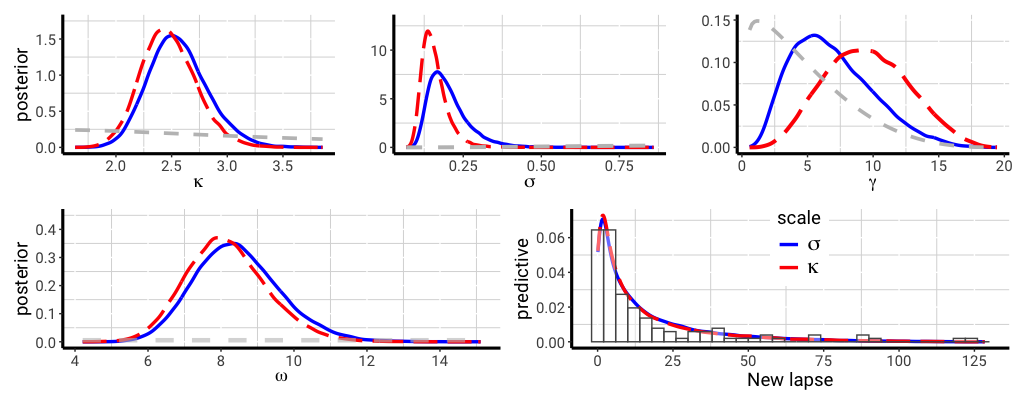}
	\caption{Posterior and posterior predictive distributions from the traffic dataset.  Priors are depicted in grey dashed lines, blue solid lines are the posteriors from the  $\sigma$ scale parameterisation and red dashed from the $\kappa$.  It is apparent that the posterior distribution of the location and the predictive distribution are not affected by the parameterisation chosen. }	\label{fig:pos_traf}
\end{figure}

There are some differences in the posterior distributions of $\sigma$ and $\gamma$, from the two parameterisations, while the posterior for $\kappa$ and $\omega$ are almost identical, as is the predictive distribution. This emphasises the identifiability issue: different combinations of $(\sigma, \gamma)$ or $(\kappa, \gamma)$ are compatible with similar values of the location, $\omega$.  Further, estimation of $\sigma$ is much more sensitive to the parameterisation, while the posterior of $\kappa$ remains stable.  In any case, the data provides strong evidence in favour of the \ekum distribution, with a log Bayes factor against $\gamma =1$, approximated with the Savage-Dickey density ratio  \citep{dickey, verdi}, of 4.2 and 13.2 from the $\sigma$ and $\kappa$ parameterisation, respectively.  The predictive probability of the next vehicle taking less than six seconds is 0.43, with (0.1, 56.92) the HPD interval of probability 0.95. Summary statistics are shown in \reta{pos_lam}.

\begin{table}[!ht]
\caption{Summary statistics from fitting the traffic time intervals data. Posterior and predictive mean, standard deviation and highest density interval (HDI) of probability 0.95, from each parameterisation.} \label{tab:pos_lam}
\centering
\begin{tabular}{r rr c rr c}
 & \multicolumn{3}{c}{$\sigma$ scale} & \multicolumn{3}{c}{$\kappa$ scale} \\
\cmidrule{2-7}
  & Mean & SD & HDI & Mean & SD & HDI \\
\midrule
$\sigma$ & 0.2 & 0.06 & $(0.104, 0.333)$ & 0.16 & 0.04 & $(0.091 , 0.241)$\\
$\kappa$ & 2.56 & 0.27 & $(2.059, 3.095)$ & 2.48 & 0.25 & $(1.993 , 2.964)$\\
$\omega$ & 8.46 & 1.18 & $(6.198,  10.779)$ & 8.16 & 1.12 & $(6.097 , 10.464)$\\
$\gamma$ & 6.91 & 3.07 & $(1.722,  12.929)$ & 9.54 & 3.17 & $(3.645 , 15.625)$\\
$t_\text{pred}$ & 15.63 & 18.72 & $(0.100 , 56.305)$ & 15.51 & 19.0 & $(0.100 , 56.750)$
\end{tabular}
\end{table}

To illustrate the key differences, we fitted the \kum distribution, with the same prior setting as the \ekum. \refi{traf_comp} overlays the corresponding posterior distributions of the location and the predictive distributions.  Note how the inclusion of the shape parameter enables the \ekum predictive distribution to retrieve the mode of the data more precisely than the \kum.  Also, the posterior distribution of the location from the \kum model is shifted to the right, the HPD interval for $\omega$ is (6.90, 12.66); with the HPD interval from the \ekum, (6.21, 10.77).

\begin{figure}[!ht]
	\centering
		\includegraphics[keepaspectratio, width=\textwidth]{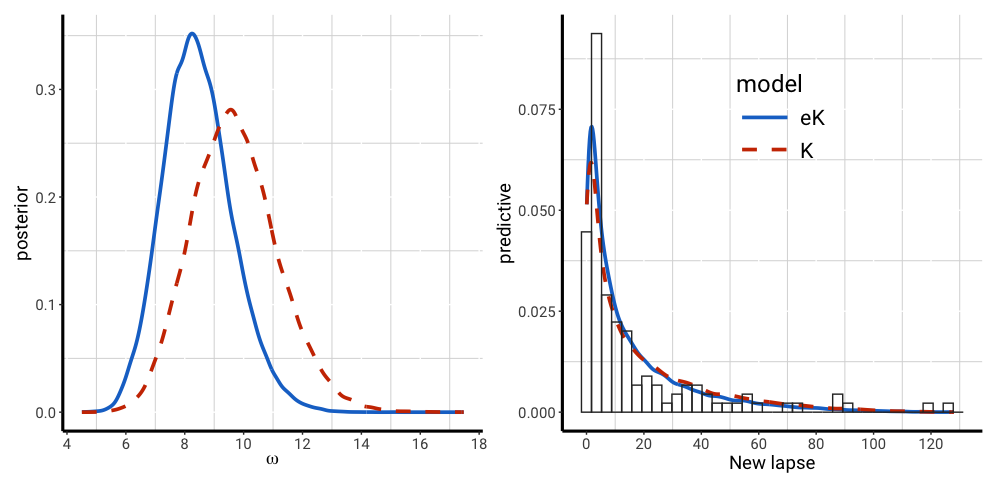}
	\caption{Traffic data. Left hand side: posterior distribution of the location.  Right had side: predictive distribution. Blue solid line shows the results from the \ekum model and the red dashed line from the \kum. The extra flexibility from the shape parameter enables the \ekum to retrieve the mode of the data more accurately. }
	\label{fig:traf_comp}
\end{figure}

Our second application involves frailty scores from adults in the UK, obtained from the Community ageing research 75+ study (CARE75+), a longitudinal study collecting data from elderly population in West Yorkshire \citep{Heavene}. Its main focus is on frailty, a clinical condition that affects an individual's ability to recover from adverse events, like illnesses or minor injuries \citep{clegg}.  Given the complexity in defining precisely frailty, there are several alternative measures \citep[see \eg][]{fried, rockwood}.  CARE75+ includes, among others, a frailty index consistent with the English longitudinal study of ageing (ELSA), with 60 deficits \citep{elsa}.  Roughly, an individual would be considered robust if their score was below 5, a score between 5 and 14 would be deemed pre-frail, and a score of 15 or above would be classified as frail; a score between 18 and 24 normally implies an elevated risk of death, requiring moving into care in most cases.   

The set we analyse is the baseline ELSA score, it comprises 1267 observations from individuals within the ages of $(78, 107)$, with 51\% females and 94\% white.  The mean and median score are 11.37 and 10, respectively, the skewness is 0.87 and the kurtosis, 3.32; a graphical summary is shown on the right hand side of \refi{dats}.

We fitted the data with the informative prior for the location and shape as in \eqref{pri_info}. The prior for the location parameter is centred at 10, with a large variance, hence \pri[\omega] = \bsc[\omega]{0.4}20{60}, and keep the prior specification as in the previous application for the remaining parameters.  The average run time was about 50 seconds for the same length and warmup as in the previous application.  Posterior and predictive distributions are shown in \refi{pos_elsa}. 

\begin{figure}[!ht]
	\centering
		\includegraphics[keepaspectratio, width = \textwidth]{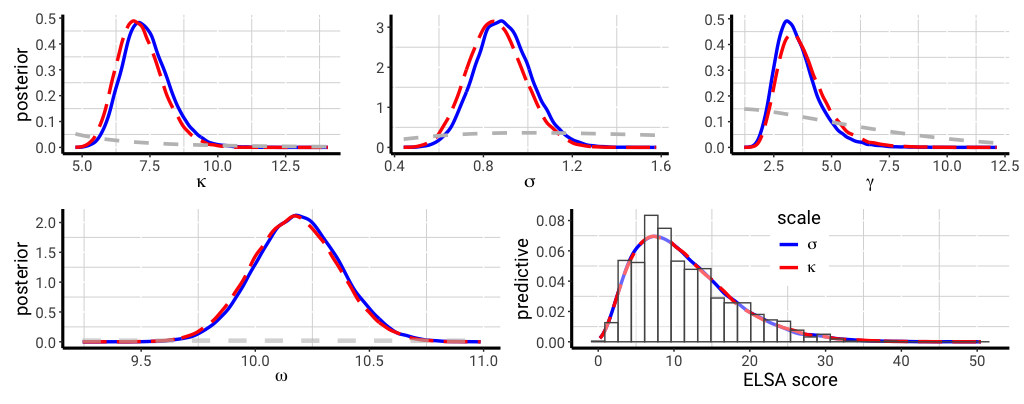}
	\caption{Posterior and posterior predictive distributions from the frailty scores dataset.  Priors are depicted in grey dashed lines, blue lines are the posteriors from the  $\sigma$ scale parameterisation and red from the $\kappa$.  It is apparent that posterior inference is not affected by the selection. }	\label{fig:pos_elsa}
\end{figure}

All posterior and predictive distributions from either parameterisation overlap significantly, as indicated by the summary statistics in \reta{pos_elsa}, due to the large sample size and regular shape of the empirical distribution. The evidence against $\gamma =1$ is overwhelming, with a log Bayes factor about 10.3. The predictive probability of a new participant in the scheme to be robust is 0.15 and 0.6 to be pre frail, in line with the observed data (0.14 and 0.61, respectively).

\begin{table}[!ht]
\caption{Summary statistics from fitting the CARE75+ ELSA frailty scores. Posterior and predictive mean, standard deviation and highest density interval (HDI) of probability 0.95, from each parameterisation.} \label{tab:pos_elsa}
\centering
\begin{tabular}{r rr c rr c}
 & \multicolumn{3}{c}{$\sigma$ scale} & \multicolumn{3}{c}{$\kappa$ scale} \\
\cmidrule{2-7}
  & Mean & SD & HDI & Mean & SD & HDI \\
\midrule
$\sigma$ & 0.89 & 0.13 & $(0.65,  1.14)$ & 0.86 & 0.13 & $(0.62, 1.11)$\\
$\kappa$ & 7.36 & 0.86 & $(5.78,  9.09)$ & 7.15 & 0.84 & $(5.59, 8.81)$\\
$\omega$ & 10.19 & 0.19 & $(9.82, 10.56)$ & 10.17 & 0.19 & $(9.80, 10.55)$\\
$\gamma$ & 3.54 & 0.96 & $(1.95,  5.46)$ & 3.77 & 1.06 & $(2.03, 5.92)$\\
$\text{ELSA}_\text{pred}$ & 11.36 & 6.45 & $(1.15, 23.93)$ & 11.36 & 6.45 & $(1.16, 24.06)$
\end{tabular}
\end{table}

\section{Closing comments} \label{sec:concs}

Dealing with data bounded in finite interval is very common in numerous fields, from scores and indices in health sciences, engineering, economics and finance, to measurements in hydrology and archeological dating.  Therefore, a vast array of alternative models and methods are available to practitioners, from monotone transformations of the data to map the support onto an unbounded space (either at one or both ends) thus enabling the use of common distributions, to using distributions that allow modelling in the original space.  The former may be the prevalent approach in practice, as it facilitates the use of off-the-shelf tools for practitioners, but carries the extra burden of appropriate interpretation of inferences, which is often not straightforward. The latter normally requires de novo fitting tools, which may deter non-experts.   

In this paper we advance the use of the exponentiated Kumaraswamy distribution for continuous BOS data, with a location-scale-shape parameterisation that removes the need for transformations and facilitates interpretability. Our Bayesian approach provides a weakly informative setting for routine use, and an informative setup to incorporate prior information with relatively little elicitation effort.  Our implementation can be easily adapted to incorporate different prior settings with small alterations to our Stan code. Additionally, our approach facilitates formal model simplification to the \kum distribution, via the Savage-Dickey density ratio; in our illustrations, however, the extra shape parameter in the \ekum is warranted. 

The model may have identifiability issues, related to the sensitivity of the \ekum to very small or very large values of the scale parameter, for a fixed median.  We address this in our Bayesian modelling by specifying a hierarchical prior that provides enough flexibility to avoid excessive shrinkage of the scale in most practical situations, and a \kum prior distribution for the shape, restricting the parameter space to (0.4, 20), which is consistent with most real applications; we also set the mode at $\gamma=1$, and let the data provide evidence against the simplification. From a practitioner standpoint, this may be a minor issue, as the key quantity for inference is typically the location and the predictive, which are insensitive to the problem. For relatively large sample sizes, however, identifiability is often not an issue, as illustrated in our applications. Given that the \ekum is more sensitive to $\sigma$, we have found in practice that fitting the $\kappa$ parameterisation is faster, requiring about 80\% of the time, compared to the $\sigma$ alternative.

\begin{small}
\bibliographystyle{BibForm} 
\bibliography{Krefs}
\end{small}
\end{document}